\documentclass[preprint, amsmath,amssymb,amssymb,aps,prl]{revtex4}
\usepackage{graphicx}
\usepackage{stmaryrd} 
\usepackage{aeguill}
\usepackage{fmtcount} 
\usepackage{dcolumn} 
\usepackage{bm}
\usepackage{braket}
\usepackage{mathrsfs}
\usepackage{times}
\usepackage{color}
\usepackage[normal]{subfigure}
\usepackage{todonotes}

\begin{document}
\title{{Coherence-Driven Topological Transition in Quantum Metamaterials}}
\author{Pankaj K. Jha$^{1}$, Michael Mrejen$^{1}$,  Jeongmin Kim$^{1}$, Chihhui Wu$^{1}$,\\ Yuan Wang$^{1}$, Yuri V. Rostovtsev$^{2}$, and Xiang Zhang$^{1,3}$}
\affiliation{$^{1}$NSF Nanoscale Science and Engineering Center(NSEC), 3112 Etcheverry Hall, University of California, Berkeley, California 94720, USA\\
$^{2}$Department of Physics, University of North Texas, Denton, Texas 76203, USA\\
$^{3}$Materials Science Division, Lawrence Berkeley National Laboratory, 1 Cyclotron Road Berkeley, California 94720, USA}
\begin{abstract}
We introduce and theoretically demonstrate a quantum metamaterial made of dense ultracold neutral atoms loaded into an inherently defect-free artificial crystal of light, immune to well-known critical challenges inevitable in conventional solid-state platforms. We demonstrate an all-optical control on ultrafast time scales over the photonic topological transition of the isofrequency contour from an open to close topology at the same frequency. This atomic lattice quantum metamaterial enables a dynamic manipulation of the decay rate branching ratio of a probe quantum emitter by more than an order of magnitude. This proposal may lead to practically lossless, tunable and topologically-reconfigurable quantum metamaterials, for single or few-photon-level applications as varied as quantum sensing, quantum information processing, and quantum simulations using metamaterials. 
\end{abstract}
\date{\today}
\maketitle
Quantum engineering of light-matter interaction at the nano-scale is an active field of research pushing the conventional boundaries in physics, material science and quantum nano-photonics and nano-engineering~\cite{Zagoskin}. In the last decade, artificial materials -metamaterials- have attracted unprecedented attention due to their exotic electromagnetic properties and may open a new realm of opportunities for quantum light-matter interaction with exciting applications~\cite{Shalaev07, Zhang11, Zheludev12,Poddubny13,Sun14}. However, harnessing metamaterials at single- or few-photon level is still an outstanding challenge owing to inevitable optical losses, and either structural or fabrication imperfections~\cite{Khurgin15}. Subsequently, the quest for quantum optical applications with metamaterial-based technologies has stimulated researchers to engineer novel lossless materials~\cite{Harry11} or construct new platforms~\cite{Quach11,Pascal14, JhaPRL15}.

In this Letter, we introduce and theoretically demonstrate a topologically reconfigurable quantum metamaterial that is immune to aforementioned severe challenges suffered by conventional metamaterial platforms. We considered dense ultracold atoms loaded into a completely off-resonant, blue-detuned, one-dimensional optical lattice~\cite{Bloch05}. We engineer the atomic response with coherent fields such that this periodic atomic density distribution displays an extreme optical anisotropy, with the dispersion contour being open, in contrast to one of close topology in natural materials. We demonstrate an all-optical quantum-coherence driven photonic topological transition in the isofrequency contour from an open response to close (ellipsoidal) at the same frequency and employ this unique control to dynamically manipulate the decay rate branching ratio of a probe quantum emitter by more than an order of magnitude. Our proposal brings together two important contemporary realms of science -cold atoms in optical lattice and metamaterials- to construct a novel architecture towards quantum metamaterials.

It is well-known that quantum coherence can drastically change the optical properties of a medium; in particular, absorption can practically vanish even at the single atom-photon level~\cite{Mucke10}. Drastic changes in the dispersion properties of a medium with excited quantum coherence have previously been theoretically and experimentally demonstrated~\cite{MOS, sau10pra, sautenkov05prl}. More recently, exotic properties such as left-handed electromagnetic responses have been theoretically proposed using dense \textit{homogenous} atomic vapors with number density $\mathcal{N}_{0}\sim 10^{23}\, m^{-3}$ or higher~\cite{Oktel04,mandel06prl,Shen06, Kastel07, yelin07prl}. However, because quantum interference is fragile, the performance of thermal atomic vapors is severely affected by collisional dephasing and Doppler broadening. These processes abbreviate the optical coherence lifetime and suppress quantum interference effects, putting an upper limit to the real part of the permittivity, Re[$\epsilon$], independent of the number density~\cite{AgarwalPRA}, and limits the figure-of-merit ($|\text{Re}[\epsilon]|/\text{Im}[\epsilon]) \sim 10^{2}$. On the other hand, with ultracold atoms in an optical lattice, these effects are negligible even at high densities, and the medium has longer coherence times~\cite{Friedman02}. Capitalizing on this one can achieve a figure-of-merit that is several orders of magnitude higher compared to the previous passive thermal atomic vapor schemes. 

Key advantages in favor of ultracold atomic lattices as a novel architecture for quantum metamaterials are the following: First, an ultrafast tunability of the topology of the optical potential and control over the atomic resonances offer a greater degree of freedom in manipulating and tuning the electromagnetic response of the atomic lattice quantum metamaterial. Second, optical lattices are rigid and inherently defect-free in sharp contrast to inevitable crystalline and fabrication defects present in conventional solid-state metamaterials.Third, completely off-resonant dipole traps also provide the freedom to precisely localize a probe atom inside the atomic lattice quantum metamaterial, and harness its exotic optical functionalities.

The schematic of atomic lattice quantum metamaterial is shown in Fig. 1 where we have considered ultracold atoms trapped by completely off-resonant and blue-detuned laser beams. We assume the spatial dependence of the atomic density grating to be Gaussian $\mathcal{N}_{\alpha}=\mathcal{N}_{0}e^{[-(z-z_{\alpha})^{2}/w^2]}$ in each period with $w$ being the $1/e$ half-width and $z_{\alpha}$ the $\alpha^{th}$ lattice site~\cite{Schilke10}. Such atomic density grating resembles the metal-dielectric multi-layered structure where the role of metal and dielectric is played by trapped atoms and vacuum respectively. Energy-level diagram of the trapped four-level atoms driven by coherent fields $\Omega_{a,d}$ and probed by a weak field $\Omega_{b}$ is shown in left-inset. Sometime, this configuration of the fields is referred to as the double dark resonance (DDR) configuration. The Hamiltonian describing the interaction between a four-level atom and the two classical fields in the rotating wave approximation is given by
\begin{equation}\label{eq1}
\begin{split}
\mathscr{H}=-\Delta_{a}|a\rangle\langle a|-(\Delta_{a}+\Delta_{b})|b\rangle\langle b|-\Delta_{d}|d\rangle\langle d|-\left(\Omega_{b}|a\rangle\langle b| +\Omega_{a}|c\rangle\langle a|+\Omega_{d}|c\rangle\langle d| +\text{h.c.}\right)
\end{split}
\end{equation} 
Here we have defined the detunings as $\Delta_{j}=\omega_{ij}-\nu_{j}$ where $\nu_{j}$ is the carrier circular frequency of the classical field. The master equation for the atom density matrix is of the Lindblad form 
\begin{equation}\label{eq2}
\frac{d\varrho}{dt}=-\frac{i}{\hbar}[\mathscr{H},\varrho]-\sum_{j}\frac{\gamma_{j}}{2}\left(\sigma^{\dagger}_{j}\sigma_{j}\varrho+ \varrho\sigma^{\dagger}_{j}\sigma_{j} -2\sigma_{j}\varrho \sigma^{\dagger}_{j}\right),
\end{equation}
in which $\sigma_{a}=|a\rangle\langle c|,  \sigma_{b}=|b\rangle\langle a|, \sigma_{d}=|d\rangle\langle c|,$ and $\gamma$ is the spontaneous decay rate on the corresponding transitions. From Eq.(\ref{eq2}), the susceptibility $\chi$ for the DDR configuration is given by~\cite{Lukin99}
\begin{equation}
\begin{split}
\chi(\Delta_{b},\Omega_{a},\Omega_{d})=i\frac{\mathcal{N}|\wp_{ab}|^{2}}{\hbar\epsilon_{0}}\frac{\Gamma_{cb}}{\Gamma_{cb}\Gamma_{ab}+|\Omega_{a}|^{2}}\left\{1+\frac{|\Omega_{a}|^{2}|\Omega_{d}|^{2}}{\Gamma_{cb}\left[\Gamma_{db}(\Gamma_{cb}\Gamma_{ab}+|\Omega_{a}|^{2})+|\Omega_{d}|^{2}\Gamma_{ab}\right]}\right\}.
\end{split}
\end{equation}
where the off-diagonal relaxation rates are given as $\Gamma_{ab}= \gamma_{ab}+i\Delta_{b}, \Gamma_{ca}=\gamma_{ca}+i\Delta_{a}, \Gamma_{cd}=\gamma_{cd}+i\Delta_{d}, \Gamma_{cb}=\gamma_{cb}+i(\Delta_{a}+\Delta_{b}), \Gamma_{ad}=\gamma_{ad}+i(\Delta_{d}-\Delta_{a}), \Gamma_{db}=i(\Delta_{d}-\Delta_{a}-\Delta_{b})$. Here $\gamma_{ab}=\gamma_{ad}=\gamma_{b}/2, \gamma_{ca}=(\gamma_{a}+\gamma_{b}+\gamma_{d})/2, \gamma_{cb}=(\gamma_{a}+\gamma_{d})/2, \gamma_{cd}=(\gamma_{a}+\gamma_{d})/2$. Within a unit cell $\alpha$ the permittivity function is given by $\epsilon_{\alpha}(z,\Delta_{b},\Omega_{a},\Omega_{d})=1+\chi_{\alpha}(z, \Delta_{b},\Omega_{a},\Omega_{d})$, where the susceptibility has a Gaussian spatial dependence given by
\begin{equation}
\begin{split}
\chi_{\alpha}(z, \Delta_{b},\Omega_{a},\Omega_{d})=\chi(\Delta_{b},\Omega_{a},\Omega_{d})e^{[-(z-z_{\alpha})^{2}/w^2]}.
\end{split}
\end{equation}
Furthermore, at high atomic density the susceptibility as modified by the local-field corrections is given by $\chi_{\text{local}}=\chi_{\alpha}/(1-\chi_{\alpha}/3).$ The drive field $(\Omega_{a})$ leads to the formation of so-called dark states that are decoupled from the coherent fields. Applying an additional weak coherent field $(\Omega_{d})$ provides more parameters to control coherences and populations in the quantum system~\cite{JhaCOP13}. We engineer the optical response of the trapped atoms with these coherent fields such that the permittivity ($\epsilon_{\alpha}$), near the atomic resonance $\omega_{ab}$, is negative at each lattice sites while the vacuum between the consecutive sites acts as a lossless dielectric of permittivity 1. We tune the magnitude of negative permittivity at each site to achieve an extreme anisotropic optical response where the atomic lattice behaves like a metal along one principal axis in contrast to dielectric nature along other orthogonal axes. This anisotropy manifest itself in the dispersion curve for extraordinary waves where the isofrequency surface is open in contrast to close topology for natural material~\cite{Comment1}.

In Fig. 2(a) we have plotted the permittivity (at $z=z_{\alpha}$) versus probe field detuning for DDR configuration. The emergence of sub-natural resonance is a signature of quantum interference~\cite{Lukin99} and its width is determined by the relaxation rates of lower levels $|b\rangle$ and $|d\rangle$ that can be very low. The Rabi frequencies and the detunings of driving fields provide efficient ways to control the position of the dip. The imaginary part of permittivity completely vanishes by carefully optimizing the drive fields $\Omega_{a}, \Omega_{d}$ and including a very weak incoherent pump on the probe transition $|b\rangle \rightarrow |a\rangle$. The inset shows the zoomed in region marked in Fig. 2(a) where we clearly see that the frequency where the imaginary part is zero, the real part is still negative. Fig. 2(b) shows the effect of the weak coherent perturbative field $\Omega_{d}$, while keeping other parameters same as Fig. 2(a), on the figure of merit (FOM) defined as $\eta=|\text{Re}[\epsilon]|/\text{Im}[\epsilon]$. Gradual increase in the FOM, with narrower bandwidth, is achieved by tuning the intensity of drive field $\Omega_{d}$. We achieved Re$[\epsilon]\sim-0.5$ at the probe frequency detuning $\Delta_{b}\sim-1.40083 \gamma_{0}$ where the FOM diverges. If we turn off the drive fields $(\Omega_{a},\Omega_{d}=0)$, four-level atom reduces to a two-level atom, we have a substantial absorption Im$[\epsilon]\sim 0.4$ and subsequently very low FOM ($\sim 4$). Thus, atoms in two-level configuration can have right dispersion, but dispersion is always accompanied by absorption, and there is no control for absorption which severely limits its applications. On the other hand with multi-level quantum emitters driven by external coherent fields one can control dispersion, practically, \textit{independent} of absorption. Typically $\gamma_{0}$ for atoms lies in the range $10^{7}-10^{8}$\,s$^{-1}$ and $\wp_{ab} \sim ea_{0}$ where $a_{0}$ is the Bohr radius. If we consider the peak density of trapped atoms $N_{0} \sim 10^{21}-10^{22}$\,m$^{-3}$ we obtain the required parameter $(\zeta/\gamma_{0} \sim 15)$ used to simulate Fig.~(2).

Next we plot the isofrequency curve $k_{z}(k_{y})$ for the transversely propagating modes (complex-valued $k_{z}$ and real-valued $k_{y})$ at the probe frequency ($\omega_{1}$) where we achieved lossless permittivity. We consider the periodicity of the atomic density grating as $\lambda_{b}/4$ where $\lambda_{b}$ is the probe wavelength. At this ratio of the grating to the probe wavelength, implementation of the effective medium theory would be inadequate to give accurate analysis for the anisotropic permittivities~\cite{Sipe12}. To simulate the isofrequency curve, we solved the wave equation as a one-dimensional eigenvalue problem using the weak form in COMSOL Multiphysics. We defined the permittivity profile over a one-dimensional segment of finite length $a$ (equal to the period) and imposed continuity boundary conditions at the extremities of the segment to simulate a slab of infinite length in the propagation direction ($z$ direction). The $z$ component of the wave-vector ($k_z$) is the complex-eigenvalue we are solving for each $k_y$ value. We are thus able to plot for the simulated system the isofrequency curve showing the $k_z$ allowed for each $k_y$.  Fig. 3(a) shows a hyperbolic-type isofrequency contour which illustrates that the effective permittivity is negative along the propagation direction ($z$-axis) but positive along the transverse direction ($x,y$-axes). Beyond the critical value of $k_{y}$, where Re$[k_{z}]$ reaches the Brillouin zone, the wave will become purely evanescent shown by the shaded grey region. The largest value of $k$ which can be harnessed is equal to $\sim 3.845/a$ where $a$ is the periodicity of the optical lattice. In our atomic lattice quantum metamaterial $a=\lambda_{b}/4$, we obtain $k_{\text{max}}\sim 4.9k_{0}$. By further detuning the trapping laser, one can decrease the periodicity and easily access higher-$k$ values. Fundamentally, upper limit to high-$k$ vectors is only limited by the periodicity of the optical potential rather than the losses and periodicity in metal-dielectric based hyperbolic metamaterial.

Metamaterials exhibiting hyperbolic dispersion have been proposed and demonstrated in several configurations, such as in arrays of metallic nanowire embedded in a dielectric host~\cite{Yao08}, multilayered metal-dielectric structure~\cite{Hoffman07, Liu07}, plasma~\cite{SZhang11,Natalia13}, metasurface~\cite{Alu15}, to name a few. These hyperbolic metamaterials have been harnessed for sub-diffraction imaging~\cite{Zubin06, Alessandro06}, negative-refraction~\cite{Hoffman07, Liu07}, photon density of states engineering~\cite{Zubin12, Menon12, Menon14}, heat transfer~\cite{Simovski13}, and so forth. However, the ability to engineer hyperbolic dispersion at will to harness propagating waves with large wavevectors has been a long sought-after goal. In that context, our atomic lattice quantum metamaterials provide a unique degree of freedom, namely an all-optical control (with weak coherent fields), on ultrafast time scales, over the photonic topological transition of the isofrequency contours. In Fig. 3(b) we have plotted the isofrequency contour, at the same probe frequency as Fig. 3a, by tuning the drive fields $\Omega_{d}, \Omega_{a}$. We see a topological transition from an open contour to a closed one, which has a profound effect on the local density of states. 

Recently, it has been proposed and theoretically demonstrated that the near-field functionality of a hyperbolic metamaterial can be matched, or even outperformed, with thin metal films~\cite{Miller15}. However, trapping and coherently addressing an isolated atom (or an array of atoms) within tens of nanometer of either a hyperbolic metamaterial (metamaterial in general) or even thin metal films, to harness its exotic near-field functionalities, is an arduous task in experiment and has been an long standing challenge~\cite{AtomChip}. In stark contrast, a completely off-resonant, blue-detuned optical lattice provide the necessary freedom to trap two-types of atoms/molecules within the same potential thus offering a solution to this outstanding challenge~\cite{Brennen99}. Furthermore, atomic lattice quantum metamaterial provides a unique opportunity to dynamically tailor and manipulate the LDOS via weak $\Omega \ll \gamma_{0}$ (coherent) external fields and coherently manipulate the decay rate branching ratio that lies at the heart of quantum optics. Such control of radiative decay rate~\cite{Agarwal,Agarwal99,Agarwal01} and their branching ratio\cite{AgarwalNJP} is not only a subject of scientific interests but has recently gained new importance owing to potential applications~\cite{Scully03, Dorfman13, DorfmanPRA11,Jha14APL,Voronine12}. To demonstrate this degree of freedom, we considered a probe atom trapped at one of the lattice sites. We assume that the transition $|1\rangle \leftrightarrow |3\rangle$ lies within the finite spectral range of engineered permittivity and $|1\rangle \leftrightarrow |2\rangle$ is far-off-resonant $\omega_{12} \gg \omega_{13}$ (for instance $|1\rangle \leftrightarrow |2\rangle$ in the near-infrared and $|1\rangle \leftrightarrow |3\rangle$ in extreme ultraviolet regime). In terms of the local coordinates the decay rate takes the form $\gamma=\gamma_{zz}$, for dipole oriented along the $z$-axis. In Fig. 4(b) we have plotted the branching ratio $\xi=\gamma_{3}/\gamma_{2}$ as a function of the coherent drive field $\Omega_{a}$ while keeping $\Omega_{d}$ and other parameters constant. Compared to two-level atom based atomic lattice $(\Omega_{a}=\Omega_{d}=0)$, we capitalize on quantum interference enabled by the weak coherent drive $\Omega_{a}$ to dynamically manipulate the decay rate branching ratio of the probe atom over an order of magnitude. Such enhancement, and reaching the critical value $\xi=1$, is imperative for steady state coherent emission in the extreme ultraviolet and X-ray regime of electromagnetic radiation~\cite{Sete12}, quantum information processing. 

The experimental implementation can be done for several atoms where the similar energy level schemes can be found, for example, in Ba, Ca, Rb, and Cs. All of them are good for trapping. In particular, we see that for Rb atoms we can choose $|d\rangle$ = 5S$_{1/2}$, $|c\rangle$ = 6P$_{3/2}$, $|b\rangle$ = 4F$_{7/2}$, $|a\rangle$ = 9D$_{3/2}$, with corresponding transitions wavelength $\lambda_d$ = 420 nm, $\lambda_b$ =1880 nm, $\lambda_a$ = 1177 nm. Similar transitions can be used for Cs as well. The number of atoms filling the lattice can be more than one, and the current available parameters of the 1D optical lattices~\cite{Zhu97, Kurn98,Lembessis05} are close to the parameters that are needed for our approach. In principle, the approach presented here can be extended to the 2D and and 3D optical lattices. Implementation of the DDR physics is both attractive and versatile, not limited to real atoms, molecules but recently extended to meta-atoms~\cite{JhaAPL13}. We anticipate our approach may bridge the gap between artificial crystals of ultracold atoms and metamaterials, and open the door for harnessing atomic lattice based quantum metamaterial at single- or few-photon-level for quantum sensing, quantum information processing, and quantum simulations.

\newpage
\begin{figure}[t!]
	{\includegraphics[scale = 0.75]{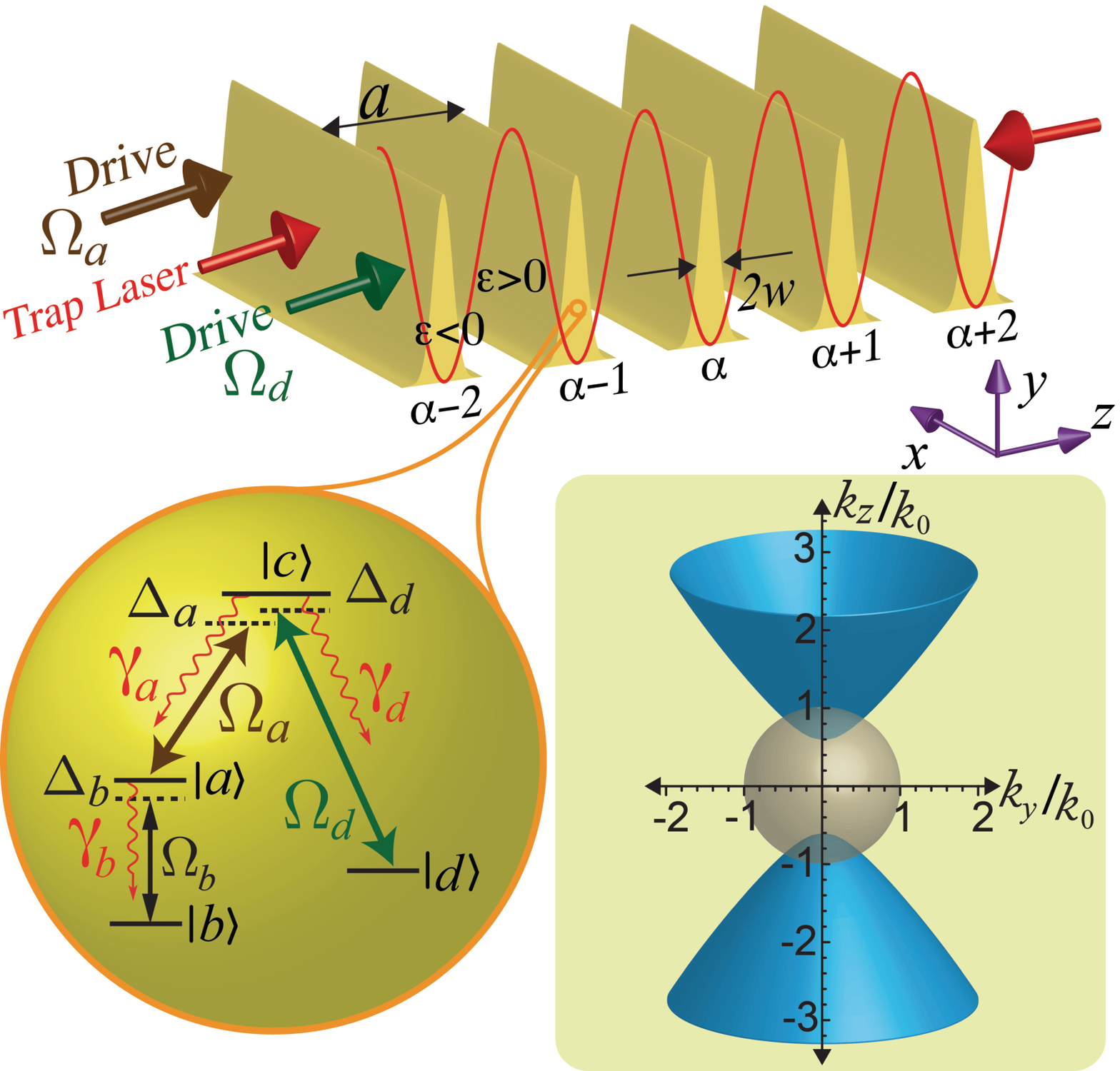}}
  \caption{Schematic of an atomic lattice quantum metamaterial: A dense ensemble of ultracold atoms loaded into the dipole traps of a one-dimensional optical lattice formed by a completely off-resonant, retro reflecting, and blue-detuned laser beams of wavelength  $\lambda_{\text{Trap}}=2a$. Electromagnetic response of the trapped atoms is engineered, using the coherent drive fields $\Omega_{a}$ and $\Omega_{d}$, to achieve negative permittivity at each lattice sites $\alpha$. The vacuum between the consecutive sites serves as a lossless dielectric. Extreme anisotropic optical response of such periodic medium, where the signs of the principal dielectric constant are different, is achieved by efficiently balancing the strength of negative permittivity at each site. For natural materials the the isofrequency contour $k_z (k_y)$ for the transversely propagating modes is closed and spherical or ellipsoidal in contrast to open and hyperbolic-type (shown in right-inset) for engineered atomic lattice.}
 \end{figure}
 \newpage
 \begin{figure}[t!]
 \centerline
	{\includegraphics[scale = 1.8]{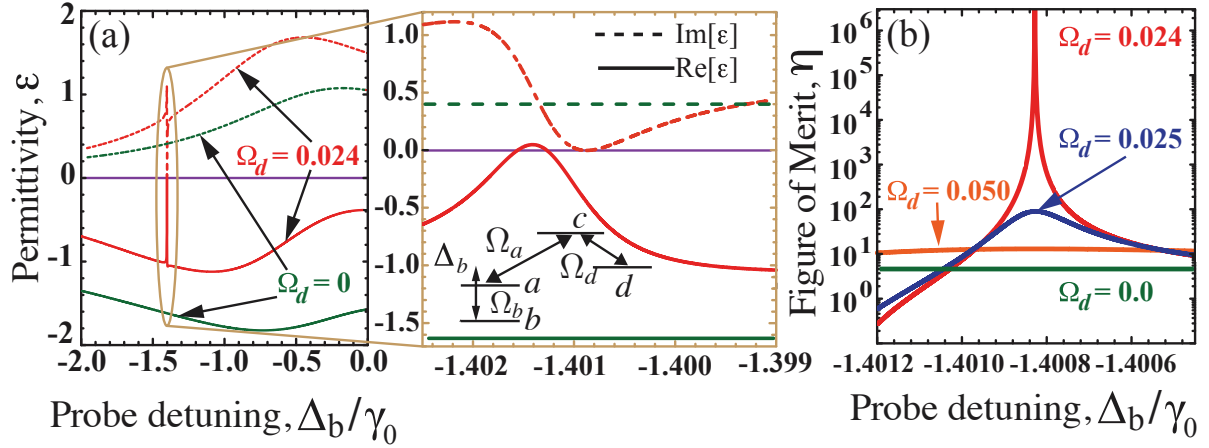}}
  \caption{Quantum-interference-induced lossless negative permittivity: (a) Plot of the permittivity against the probe detuning $\Delta_{b}$. We achieved complete cancellation of absorption (Im$[\epsilon]=0$) at the probe detuning $\Delta_{b}=-1.40083\gamma_{0}$ by incorporating a very weak incoherent pump on the $|b\rangle \rightarrow |a\rangle$ transition. At this detuning the real part is Re$[\epsilon]\sim-0.5$. The inset is a zoomed in region of (a) that shows the frequency ($\omega_{1}$), where the imaginary part of the permittivity goes to zero, the real part is negative. Without the drive field we have a significantly absorption (Im($\epsilon$) = 0.4). (b) Plot of the figure of merit defined as $\eta=|\text{Re}[\epsilon]|/\text{Im}[\epsilon]$ against the probe detuning for different values of the weak coherent perturbative field $\Omega_{d}$. By fine tuning of the field $\Omega_{d}$, FOM increases and eventually diverges. Parameters for numerical simulations are $\gamma_{b}=0.62, \gamma_{a}=0.86, \gamma_{d}=1.092, \Omega_{a}=1.3, \Omega_{d}=0.024, \Delta_{d}=-1.40073$. Rabi frequency and decay rate has been normalized to $\gamma_{0}$}
 \end{figure}
 \newpage
 \begin{figure}[t!]
\centerline
	{\includegraphics[scale = 1.95]{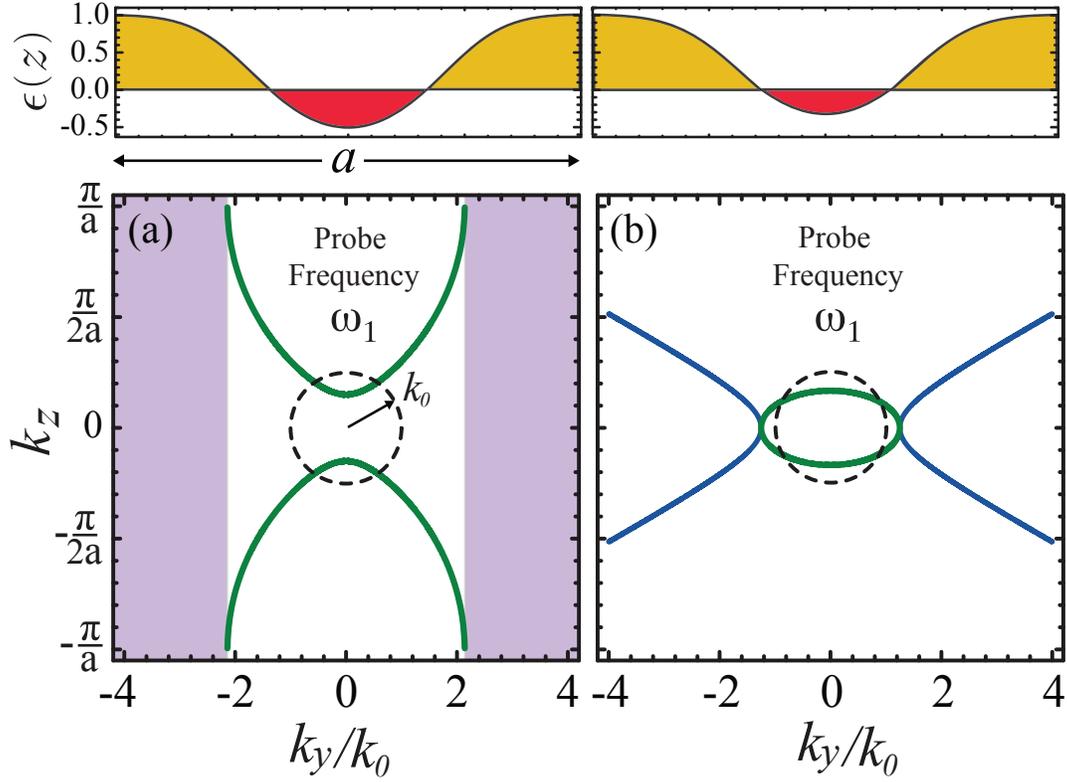}}
  \caption{Quantum-coherence driven topological transition of the iso-frequency contour at the same frequency. Iso-frequency contours $k_{z}(k_{y})$ for the transversely propagating electromagnetic modes (complex-valued $k_{z}$ and real-valued $k_{y})$ at the probe frequency ($\omega_{1}$) where Im$(\epsilon \sim 0)$. By optimizing the coherent optical fields $\Omega_{a}, \Omega_{d}$ we achieved both (a) hyperbolic-type and (b) elliptical contours at the same frequency. The dashed black lines in (a) and (b) indicates the free-space IFC. Spatial variation of the permittivity $\epsilon(z)$ within a unit cell corresponding to the IFC (a) and (b) are shown in the upper panels respectively. For numerical simulations we considered $\lambda_{\text{Trap}}=2.5d= 0.25\lambda_{b}$, (a) $\Omega_{a}=1.3, \Omega_{d}=0.024$ and (b) $\Omega_{a}=1.15, \Omega_{d}=0.0189$. Rabi frequency and decay rate has been normalized to $\gamma_{0}$.}
 \end{figure}
  \newpage
 \begin{figure}[t!]
\centerline
	{\includegraphics[scale = 1.1]{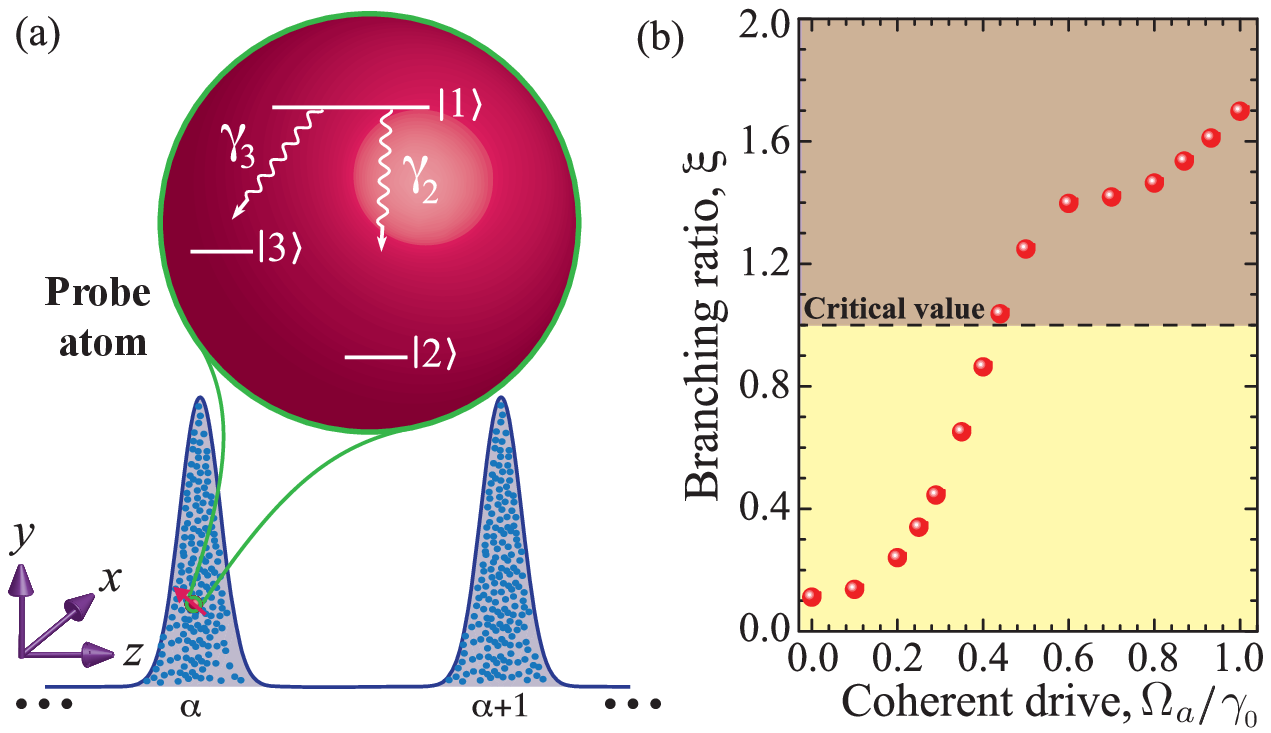}}
  \caption{Optical control of a quantum emitter decay rate branching ratio. (a) Schematic of a probe quantum emitter (three-level atom in $\Lambda$-configuration) trapped at one of the lattice sites. (b) Plot of the branching ratio defined as $\xi=\gamma_{3}/\gamma_{2}$ for different values of the drive field $\Omega_{a}$. We achieve the critical value of $\xi=1$ using a weak a drive field $\Omega_{a} \sim 0.5 \gamma_{0}$. For numerical simulations the other parameters are same as Fig. (2).}
 \end{figure}

\begin{thebibliography}{99}
\bibitem{Zagoskin} A. M. Zagoskin, \textit{Quantum Engineering: Theory and Design of Quantum Coherent Structures}, (Cambridge University Press, England, 2011)
\bibitem{Shalaev07}V. Shalaev, Nature Photonics \textbf{1}, 41(2007).
\bibitem{Zhang11}Y. Liu and X. Zhang, Chem. Soc. Rev. \textbf{40}, 2494(2011).
\bibitem{Zheludev12} N. I. Zheludev and Y. S. Kivshar, Nat. Mater. \textbf{11}, 917 (2012).
\bibitem{Poddubny13} A. Poddubny, I. Iprsh, P. Belov, and Y. Kivshar, Nature Photonics \textbf{7}, 948(2013).
\bibitem{Sun14}J. Sun, N. M. Litchinitser, and J. Zhour, ACS Photonics 1, 293(2014).
\bibitem{Khurgin15}J. B. Khurgin, Nat. Nano. \textbf{10}, 2 (2015).
\bibitem{Harry11} A. Boltasseva and H. A. Atwater, Science \textbf{331}, 290 (2011).
\bibitem{Quach11}J. Q. Quach, C-H Su, A. M. Martin, A. D. Greentree, and L. C. L. Hollenberg, Opt. Express \textbf{19}, 11018(2011). 
\bibitem{Pascal14} P. Macha, G. Oelsner,	J-M. Reiner, M. Marthaler, S Andre, G. Schon,	U. Hubner, H-G. Meyer, E. Ilichev, and A. V. Ustinov, Nature Comm. \textbf{5}, 6146(2014).
\bibitem{JhaPRL15}P. K. Jha, X. Ni, C. Wu, Y. Wang and X. Zhang, Phys. Rev. Lett. \textbf{115}, 025501(2015).
\bibitem{Bloch05} I. Bloch, Nature Physics \textbf{1}, 23(2005).
\bibitem{Mucke10} M. Mucke,	E. Figueroa, J. Bochmann, C. Hahn,	K. Murr, S. Ritter, C. J. Villas-Boas, and G. Rempe, Nature \textbf{465}, 755(2010).
\bibitem{MOS} M. O. Scully, and M. S. Zubairy, \textit{Quantum Optics}, (Cambridge University Press, Cambridge, England, 1997).
\bibitem{sau10pra} V. A. Sautenkov, H. Li, Y. V. Rostovtsev, M. O. Scully, Phys. Rev. A \textbf{81}, 063824 (2010).
\bibitem{sautenkov05prl} V. A. Sautenkov, Y. V. Rostovtsev, H. Chen, P. Hsu, Girish S. Agarwal, and M. O. Scully, Phys. Rev. Lett. \textbf{94}, 233601 (2005).
\bibitem{Oktel04}M. O. Oktel, and O. E. Mustecaplioglu, Phys. Rev. A \textbf{70}, 053806(2004). 
\bibitem{mandel06prl} Q. Thommen and P. Mandel, Phys. Rev. Lett. \textbf{96}, 053601 (2006).
\bibitem{Shen06}J. Q. Shen, J. Mod. Opt. \textbf{53}, 2195(2006).
\bibitem{Kastel07} J. Kastel, M. Fleischhauer, Phys. Rev. Lett. \textbf{98}, 069301(2007). 
\bibitem{yelin07prl} J. Kastel, M. Fleischhauer, S.F. Yelin, and R.L. Walsworth, Phys. Rev. Lett. \textbf{99}, 073602 (2007). 
\bibitem{AgarwalPRA} G. S. Agarwal and R. W. Boyd, Phys. Rev. A \textbf{60} R2681(1999).
\bibitem{Friedman02}N. Friedman, A. Kaplan, and N. Davidson, Adv. At. Mol. Opt. Phys. \textbf{48}, 99(2002).
\bibitem{Schilke10}A. Schilke, C. Zimmermann, and W. Guerin Phys. Rev. A \textbf{86}, 023809(2012).
 \bibitem{Lukin99} M. D. Lukin, S. F. Yelin, M. Fleischhauer, and M. O. Scully, Phys. Rev. A \textbf{60}, 3225 (1999).  
 \bibitem{JhaCOP13} P. K. Jha, Coherent Opt. Phenom. \textbf{1}, 25(2013).  
  \bibitem{Comment1} By carefully optimizing the optical parameters we also achieve Type II response. See Fig. (1) of the Supplementary material. Furthermore,  by further tuning the resonance and the coherent drive field strength we also achieved lossless epsilon-near-zero response (see Fig. S2 in Supplementary Material). 
 \bibitem{Sipe12} O. Kidwai, S. V. Zhukovsky, and J. E. Sipe, Phys. Rev. A \textbf{85}, 053842(2012).
 \bibitem{Yao08} J. Yao, Z. Liu, Y. Liu, Y. Wang, C. Sun, G. Bartal, A. M. Stacy, X. Zhang, Science \textbf{321}, 930 (2008).
\bibitem{Hoffman07}A. J. Hoffman, L. Alekseyev, S. S. Howard, K. J. Franz, D. Wasserman, V. A. Podolskiy, E. E. Narimanov, D. L. Sivco, and C. Gmachl, Nat. Mater. \textbf{6}, 946(2007).
\bibitem{Liu07} Z. Liu, H. Lee, Y. Xiong, C. Sun, X. Zhang, Science \textbf{315}, 1686 (2007).
\bibitem{SZhang11} S. Zhang, Y. Xiong, G. Bartal, X. Yin and X. Zhang, Phys. Rev. Lett. \textbf{106}, 243901(2011).
\bibitem{Natalia13} Z. A. Kudyshev, M. C. Richardson and N. M. Litchinister, Nature Comm. \textbf{4}, 3557(2013).
\bibitem{Alu15}J. S. Gomez-Diaz, M. Tymchenko, and A. Alu, Phys. Rev. Lett. \textbf{114}, 233901(2015).
\bibitem{Zubin06} Z. Jacob, L. V. Alekseyev, E. Narimanov, Opt. Express \textbf{14}, 8247(2006).
\bibitem{Alessandro06} A. Salandrino, N. Engheta, Phys. Rev. B \textbf{74}, 075103 (2006).
\bibitem{Zubin12} Z. Jacob, I. I. Smolyaninov, and E. E. Narimanov, Appl. Phys. Lett. \textbf{100}, 181105(2012).
\bibitem{Menon12}H. N. S. Krishnamoorthy, Z. Jacob, E. Narimanov, I. Kretzschmar, and V. M. Menon, Science \textbf{336}, 205 (2012).
 \bibitem{Menon14} H. N. S. Krishnamoorthy, Y. Zhou, S. Ramanathan, E. Narimanov, and V. M. Menon, App. Phys. Lett. \textbf{104}, 121101(2014).
\bibitem{Simovski13}C. Simovski, S. Maslovski, I. Nefedov, and S. Tretyakov,  Opt. Express \textbf{21} 14988(2013).
\bibitem{Miller15} O. D. Miller, S. G. Johnson, and A. W. Rodriguez, Phys. Rev. Lett. \textbf{112}, 157402(2014).
\bibitem{AtomChip}J. Reichel and V. Vuletic, editors, \textit{Atom Chips},  Wiley-VCH, Weinheim, Germany, 2011.
 \bibitem{Brennen99} G. K. Brennen, C. M. Caves, P. S. Jessen, and I. H. Deutsch, Phys. Rev. Lett. \textbf{82}, 1060(1999).
\bibitem{Agarwal} G. S. Agarwal, \textit{Quantum Statistical Theories of Spontaneous Emission and Their Relation to Other Approaches, Springer Tracts in Modern Physics: Quantum Optics} Vol. 70, edited by G. Mohler et al. (Springer-Verlag, Berlin, 1974).
\bibitem{Agarwal99} G. S. Agarwal Phys. Rev. A \textbf{61}, 013809 (1999).
\bibitem{Agarwal01} G. S. Agarwal, M. O. Scully, and H. Walther, Phys. Rev. Lett. \textbf{86}, 4271 (2001).
\bibitem{AgarwalNJP} G. S. Agarwal and S. Das, New Journal of Physics \textbf{10}, 013014(2008).
\bibitem{Scully03} M. O. Scully, M. Suhail Zubairy, G. S. Agarwal, and H.Walther, Science \textbf{299}, 862(2003).
\bibitem{Dorfman13}K. E. Dorfman, P. K. Jha, D. V. Voronine, P. Genevet, F. Capasso, and M. O. Scully, Phys. Rev. Lett. \textbf{111}, 043601(2013).
\bibitem{DorfmanPRA11}K. E. Dorfman, P. K. Jha, and S. Das, Phys. Rev. A \textbf{84}, 053803(2011). 
 \bibitem{Jha14APL}P. K. Jha, M. Mrejen, J. Kim, C. Wu, X. Yin, Y. Wang, X. Zhang, Appl. Phys. Lett. \textbf{105}, 111109 (2014).
 \bibitem{Voronine12} D. V. Voronine, A. M. Sinyukov, X. Hua, K. Wang, P. K. Jha, E. Munusamy, S. E. Wheeler, G. R. Welch, A. V. Sokolov, and M. O. Scully, Sci. Rep. \textbf{2}, 891(2012).
\bibitem{Sete12}  E. A. Sete, A. A. Svidzinsky, Y. V. Rostovtsev, H. Eleuch, P. K. Jha, S. Suckewer, and M. O. Scully, IEEE J. Sel. Top. Quantum Electron. \textbf{18}, 541 (2012).
\bibitem{Zhu97} X. D. Zhu Opt. Lett. \textbf{22},1890(1997).
 \bibitem{Kurn98}D. M. Stamper-Kurn, M. R. Andrews, A. P. Chikkatur, S. Inouye, H.-J. Miesner, J. Stenger, and W. Ketterle, Phys. Rev. Lett. \textbf{80}, 2027(1998).
\bibitem{Lembessis05}V. E. Lembessis and D. Ellinas, J. Opt. B: Quantum Semiclass. Opt. \textbf{7} 319 (2005). 
 \bibitem{JhaAPL13} P. K. Jha, X. Yin and X. Zhang, Appl. Phys. Lett. \textbf{102}, 091111 (2013).
\end{thebibliography}
 \end{document}